# The $^{13}$C-pockets in AGB Stars and Their Fingerprints in Mainstream SiC Grains

**Nan Liu**[a,b], **Andrew M. Davis**[a,b,c], **Roberto Gallino**[d], **Michael R. Savina**[b,e], **Sara Bisterzo**[d], **Frank Gyngard**[f], **Michael J. Pellin**[a,b,c,e], **Nicolas Dauphas**[a,b,c]

[a]*Department of the Geophysical Sciences,* [b]*Chicago Center for Cosmochemistry,* [c]*Enrico Fermi Institute, The University of Chicago*
*5734 S. Ellis Ave., Chicago, IL 60637, USA*

[d]*Dipartimento di Fisica*
*Università di Torino*
*Torino I-10125, Italy*

[e]*Materials Science Division*
*Argonne National Laboratory*
*Argonne, IL 60439, USA*

[f]*Laboratory for Space Sciences*
*Washington University, St. Louis*
*St. Louis, MO 63130, USA*

E-mails: lnsmile@uchicago.edu; a-davis@uchicago.edu; gallino@ph.unito.it; savina@anl.gov; bisterzo@ph.unito.it; fgyngard@wustl.edu; pellin@anl.gov; dauphas@uchicago.edu

We identify three isotopic tracers that can be used to constrain the $^{13}$C-pocket and show the correlated isotopic ratios of Sr and Ba in single mainstream presolar SiC grains. These newly measured data can be explained by postprocess AGB model calculations with large $^{13}$C-pockets with a range of relatively low $^{13}$C concentrations, which may suggest that multiple mixing processes contributed to the $^{13}$C-pocket formation in parent AGB stars.

*XIII Nuclei in the Cosmos*
*7-11 July, 2014*
*Debrecen, Hungary*

## 1. Introduction

AGB stars are the stellar site for the main *s*-process (slow neutron capture process) with $^{13}$C being the major neutron source, via $^{13}$C($\alpha$,n)$^{16}$O [1]. Formation of $^{13}$C via $^{12}$C($p,\gamma$)$^{13}$N($\beta^+\nu$)$^{13}$C within the so-called "$^{13}$C-pocket", a region beneath the bottom of the convective envelope, requires mixing protons from the convective H-rich envelope into the radiative He-rich intershell, a local unstable thermodynamic situation that is extremely difficult to treat in stellar codes. Many physical mechanisms may compete for the pocket formation, including overshoot, gravity waves, magnetic buoyancy, and Eddington-Sweet (ES) and Goldreich-Schubert-Fricke (GSF) instabilities in rotating AGB stars [2−5]. Currently, it is unclear which process(es) are responsible for the $^{13}$C-pocket formation. Thus, the $^{13}$C neutron source in AGB stellar model calculations suffers from uncertainties in the adopted $^{13}$C profile, $^{13}$C concentration, and $^{13}$C-pocket mass [6].

Previous constraints on the $^{13}$C-pocket were mainly derived based on the solar system *s*-process pattern and spectroscopic observations of [hs/ls] and [Pb/hs] ratios in stars with different metallicities. However, the solar *s*-process pattern is the result of nucleosynthesis in all previous generations of AGB stars, so its quantitative explanation suffers from additional uncertainties associated with Galactic Chemical Evolution (GCE) assumptions. In addition, spectroscopic observations of AGB stars mainly provide information about elemental abundances, while nuclide abundance (isotope abundance) distributions in individual stars are required to obtain detailed constraints on the $^{13}$C-pockets.

Presolar grains are ancient stellar relics that survived destruction in the interstellar medium and the early solar system, and were discovered in pristine meteorites. Based on extensive studies of isotopic anomalies, a majority of presolar SiC grains (mainstream grains) came from low-mass AGB stars, the stellar site for *s*-process. The mainstream grains contain *s*-process isotopic signatures inherited from their parent AGB stars. In previous heavy element isotopic measurements, presolar grains were contaminated with solar system material, which obfuscated data interpretation [6]. Also, most of previous studies measured isotopic ratios of only one heavy element and are insufficient to provide stringent constraints on the $^{13}$C-pocket. In order to systematically investigate the $^{13}$C-pocket uncertainties, we acid-cleaned presolar SiC grains after their separation from the bulk Murchison meteorite. In addition, we identified three sensitive isotopic tracers ($^{88}$Sr/$^{86}$Sr, $^{92}$Zr/$^{94}$Zr, and $^{138}$Ba/$^{136}$Ba) and simultaneously measured Sr and Ba isotopic compositions in these single acid-cleaned SiC grains using resonance ionization mass spectrometry. The analyzed grains were later on identified as mainstream by measuring C and Si isotope ratios in the remainder of each grain, using a NanoSIMS.

## 2. Torino Postprocess AGB Model

A parameterized $^{13}$C-pocket is adopted in Torino postprocess model calculations in order to systematically investigate the $^{13}$C-pocket uncertainties and to provide constraints on the parameters [6,7]. The $^{13}$C-pocket is characterized by the following three variables and assumed to stay constant during each interpulse in the model calculations:

(1) $^{13}$C mass fraction within the pocket, $X(^{13}C)$. The $X(^{13}C)$ values in the ST(standard) *case*, in which the solar *s*-process pattern is well reproduced with a 0.5 $Z_\odot$ AGB star, are divided by the corresponding factors in case names from D7.5 to D1.3. Such a range could be caused by variation in the rotation speeds of AGB stars [5].

(2) $^{13}$C profile, *decreasing-with-depth or flat*. We started with a decreasing-with-depth $^{13}$C-pocket (Three-zone models). It is subdivided into three different zones: I, II and III, which are numbered in order of increasing stellar radius. $X(^{13}C)$ decreases from Zone-III to Zone-I with a fixed slope while remaining constant within each zone. The flat $^{13}$C-pocket (Zone-II models) consists of Zone-II only, which could be a result of smoothing by the GSF instability in rotating AGB stars [5].

(3) $^{13}$C-pocket mass, $M(^{13}C)$. The $^{13}$C-pocket mass in Three-zone or Zone-II model calculations is multiplied (p) or divided (d) by the numbers in corresponding model names (*e.g.*,

Zone-II_p2). The pocket mass ranges from $(4-80)\times10^{-4}$ $M_\odot$. An enlarged pocket could result from the overshoot process plus the ES instability in rotating AGB stars and/or magnetic buoyancy [4,5].

## 3. Sensitive Isotopic Tracers

In the classical approach for *s*-process nucleosynthesis calculations, equilibrium in the neutron-capture flow is obtained ($dN_A/dt=0$) in the mass regions between neutron-magic numbers ($N$=50, 82, 126) [8]. Thus, to first-order, in mass regions between neutron-magic numbers, the abundance of a nuclide made in the *s*-process is inversely proportional to its neutron-capture cross section at relevant stellar conditions. Consequently, the abundances of a number of *s*-process nuclei, especially their relative abundances, are barely affected by AGB stellar conditions. Their abundances are therefore not very useful in constraining the uncertainties in the $^{13}$C-pocket. A few exceptions, however, exist along the *s*-process path.

### *3.1. s-Process Bottlenecks.*

The bottleneck effect is caused by the small Maxwellian-Averaged Cross Sections (MACSs) of neutron-magic nuclei. The MACSs of $^{88}$Sr and $^{138}$Ba are a factor of ten lower than those of their neighboring even-numbered isotopes, $^{86}$Sr and $^{136}$Ba. Only a small fraction of the *s*-process neutron flow starting at the iron peak can pass through these bottlenecks. Thus, the values of $^{88}$Sr/$^{86}$Sr and $^{138}$Ba/$^{136}$Ba are sensitive to the adopted $^{13}$C-pockets in AGB model calculations.

### *3.2. Temperature Dependence of MACS.*

To a first approximation, the MACSs of most nuclei follow the $1/v_T$ (Maxwellian-averaged velocity at temperature *T*) rule [9]. However, the $^{92}$Zr MACS deviates from the rule by 30%, while the $^{94}$Zr MACS strictly follows the rule from 8 to 23 keV (AGB neutron energy regime). The relative production ratio of $^{92}$Zr and $^{94}$Zr therefore varies with temperature and, in turn, depends on the neutron sources, $^{13}$C and $^{22}$Ne ($^{22}$Ne is a minor neutron source in AGB stars), which provide neutrons to the *s*-process at different stellar temperatures. The predicted $^{92}$Zr/$^{94}$Zr value therefore varies with the $^{13}$C-pocket adopted in AGB model calculations [7].

## 4. Constraints on the $^{13}$C-pocket

As shown in Figure 1, the isotopic compositions of mainstream SiC grains in this study define a tight cluster (grain-concentrated region), with δ($^{88}$Sr/$^{86}$Sr) ranging from −200‰ to 0‰ and δ($^{138}$Ba/$^{136}$Ba) from −400‰ to −200‰. From Figure 1, we can see that for the three variables, the order of importance in explaining the variation of isotopic compositions in mainstream SiC grains is $^{13}$C mass fraction, $^{13}$C-pocket mass, and $^{13}$C profile within the $^{13}$C-pocket. In addition, lower-than-D2 cases play important roles in matching the grain-concentrated region.

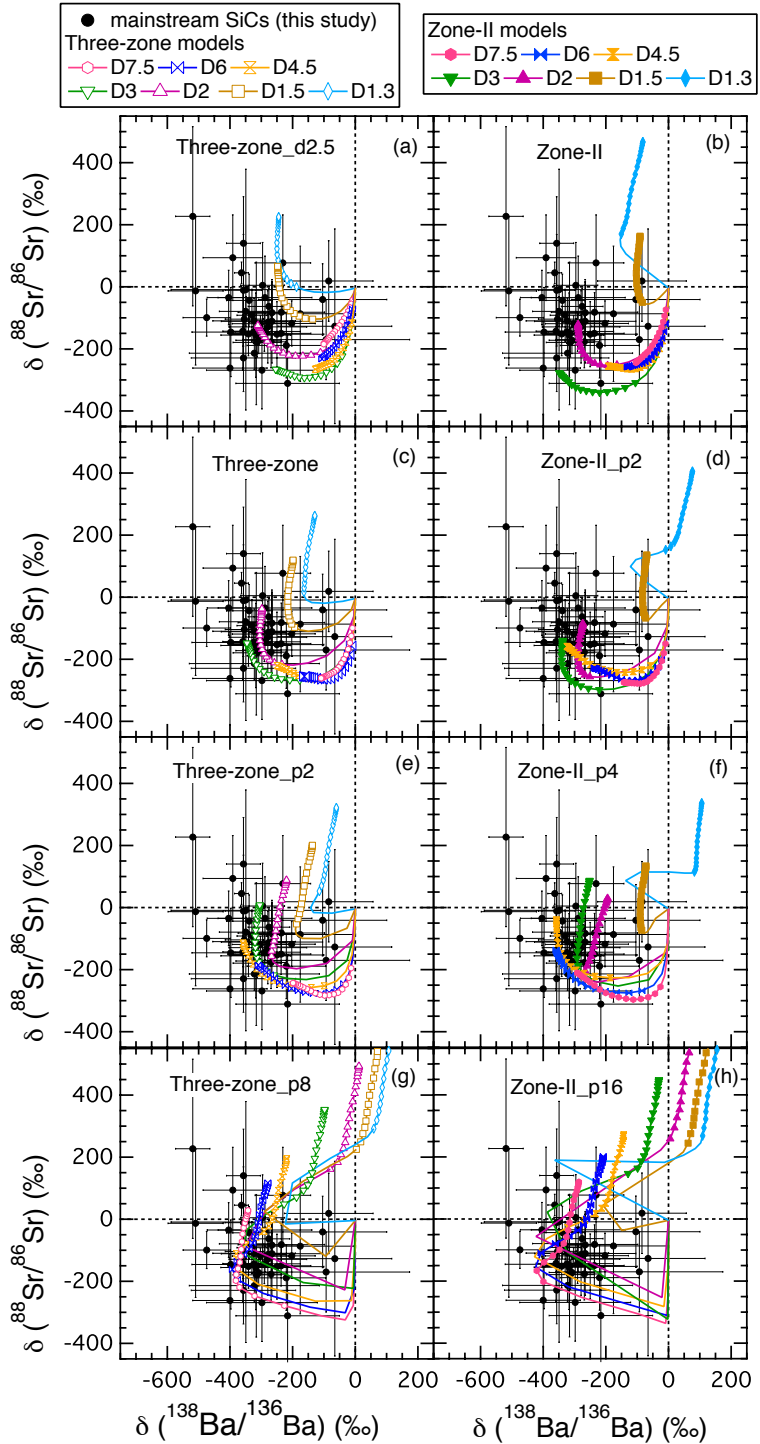

Figure 1. Mainstream SiC grain data [10] are shown as black dots with 2σ errors. The delta values are defined as

$$\delta\left(\frac{^{i}Sr}{^{86}Sr}\right) = \left[\frac{\left(\frac{^{i}Sr}{^{86}Sr}\right)_g}{\left(\frac{^{i}Sr}{^{86}Sr}\right)_{std}} - 1\right] \times 1000$$

(*g* stands for grain and *std* for standard). Each Torino AGB model predictions evolve from zero (initial stellar composition) along colored lines, with symbols plotted on top of the lines when C/O>1 in the envelope. As SiC is a highly reduced mineral, it can only condense in a carbon-rich environment [11]. Thus, the grain data can only be compared to the model predictions shown as colored lines with symbols. In each plot, different colors correspond to different ¹³C mass fractions (¹³C concentrations). Each column corresponds to a different ¹³C profile (decreasing-with-depth versus flat), and each row corresponds to a different pocket mass from $(4-80) \times 10^{-4}$ $M_\odot$.

Model predictions with even lower-mass $^{13}$C-pockets ($^{13}$C-pocket mass<$5\times10^{-4}$ $M_\odot$) are not shown, because predictions for Sr and Ba isotope ratios barely change below this pocket mass except those for δ($^{84}$Sr/$^{86}$Sr), which continue to increase with decreasing pocket mass until $1.3\times10^{-4}$ $M_\odot$. On the other hand, because the $^{13}$C-pocket forms within the He-intershell, the pocket mass must be less than the He-intershell mass, which decreases from $1\times10^{-2}$ to $8\times10^{-3}$ $M_\odot$ from the first to last TP during the AGB phase [1]. In addition, as the pocket mass is constant for different TPs in Torino AGB models, the maximum $^{13}$C-pocket mass allowed in Torino AGB models is therefore $8\times10^{-3}$ $M_\odot$ (Figures 1g,h). In Figures 1g,h, Three-zone and Zone-II model predictions above the D4.5 case are shifted to both higher δ($^{88}$Sr/$^{86}$Sr) and δ($^{138}$Ba/$^{136}$Ba) values, while calculations below D4.5 show coiled shapes, avoiding the grain-concentrated region. Thus, only model calculations with lower-than-D4.5 mass fractions in Figures 1g,h can reach the grain regime during carbon-rich phases.

Based on Figure 1, two first-order conclusions can be drawn:

(1) Varying $^{13}$C-pockets with pocket mass < $1\times10^{-3}$ $M_\odot$ are not sufficient to cover the whole range of the grain data; higher-mass $^{13}$C-pockets ($\geq 1\times10^{-3}$ $M_\odot$) are required;

(2) In order to reach the grain data region, models must adopt inversely correlated pocket mass and $^{13}$C mass fraction; *e.g.,* models with higher pocket masses require lower $^{13}$C mass fractions.

## 5. Implications for $^{13}$C-pocket Formation

The correlated $^{88}$Sr/$^{86}$Sr and $^{138}$Ba/$^{136}$Ba ratios in mainstream SiC grains point towards the existence of large $^{13}$C-pockets with a range of relatively low $^{13}$C concentrations ($^{13}$C mass fractions) in AGB stars. As convective overshoot is not expected to be a stochastic process [12], the range of $^{13}$C concentrations among parent AGB stars of mainstream SiC grains could more likely be explained by different angular velocity gradients at the core-envelope interface in rotating AGB stars with different initial rotation velocities possibly interacting with the generated magnetic fields. Higher pocket mass than that produced by overshoot (< $1\times10^{-3}$ $M_\odot$) could be obtained in model calculations by considering other processes, such as gravity waves excited by turbulent motions near the base of the convective envelope [3] or magnetic fields [4]. In fact, it cannot be excluded that various mixing processes may operate simultaneously, interacting with each other. For instance, gravity waves are expected to work during TDU episodes and magnetic fields develop in rotating stars. One or more other mixing processes in addition to overshoot and rotation may be invoked to shape the large $^{13}$C-pockets with low $^{13}$C concentrations suggested by the grain data from this study.